\documentclass[%
reprint,
superscriptaddress,
amssymb,
prl,
]
{revtex4-1}

\usepackage[english]{babel} 
\usepackage [table]{xcolor}

\usepackage{graphicx} 

\usepackage{amsmath}
\usepackage{amsthm}
\usepackage{amsfonts}
\usepackage{color}

\usepackage{float}
\usepackage{textcomp}
\usepackage{graphicx}
\usepackage{dcolumn}
\usepackage{bm}
\usepackage{hyperref}
\usepackage[mathlines]{lineno}

\begin{document}



\title {Analysis of high quality superconducting resonators: consequences for TLS properties in amorphous oxides}

\author{J.~Burnett}

\affiliation{London Centre for Nanotechnology, University College London, 17-19 Gordon Street, WC1H 0AH, UK}

\author{L. Faoro}

\affiliation{Laboratoire de Physique Theorique et Hautes Energies, CNRS UMR 7589,
Universites Paris 6 et 7, 4 place Jussieu, 75252 Paris, Cedex 05,
France}

\affiliation{Department of Physics and Astronomy, Rutgers The State University
of New Jersey, 136 Frelinghuysen Road, Piscataway, 08854 New Jersey,
USA}

\author{T.~Lindstr\"{o}m}

\affiliation{ National Physical Laboratory, Hampton Road, Teddington, TW11 0LW,
UK}

\email{tobias.lindstrom@npl.co.uk}

\selectlanguage{english}

\date{\today}

\begin{abstract}
$1/f$ noise caused by microscopic Two-Level Systems (TLS) is known to be very detrimental to the performance of superconducting quantum devices but the nature of these TLS is still poorly understood. Recent experiments with superconducting resonators indicates that interaction between TLS  in the oxide at the film-substrate interface is  not negligible. Here we present data on the loss and  $1/f$ frequency noise from  two different Nb resonators with and without Pt capping and discuss what conclusions can be drawn regarding the  properties of TLS  in amorphous oxides. We also estimate the concentration and dipole moment of the TLS.\end{abstract}

\maketitle

\section{Introduction}
Superconducting electronics has become a frontrunner in the race to create viable applications of  solid state quantum technology. For many of these devices superconducting resonators play a fundamental role, both as an integral part of quantum circuits and as a test-bed for developing fabrication technology. Recently, planar on-chip superconducting resonators with internal quality factors $Q_{i}$ above 10$^6$ have been developed \cite{megrant2012planar,bruno2015reducing}. The primary challenge in their development has been in the understanding and mitigation of parasitic Two Level Systems (TLS) which lead to a decrease in $Q_{i}$ in these  resonators at mK temperatures and single photon energies where superconducting qubits are operated\cite{lindstrom2009properties,macha2010losses}. The presence of TLS is also known to be directly detrimental to coherence times of superconducting qubits. 

Despite a large research effort and improvements in quality factors there is as yet no method of completely eliminating parasitic TLS. Instead, the community has found ways of circumventing the problem by using 3D cavities\cite{paik2011observation}. However, planar devices will almost certainly be necessary in future large-scale integrated quantum circuits meaning  the TLS problem will nevertheless have to be solved.  Hence, a better understanding of the nature of these TLS  is crucial. 

Historically, the so-called Standard Tunnelling Model (STM)  \cite{anderson1972anomalous,phillips1972tunneling} -  first developed to study amorphous glasses in the 1970s - has been used to model the effect of TLS on superconducting resonators. The STM assumes that the TLS have a uniform distribution of the energy splitting and that the interactions between TLS are negligible. 
Observation of temperature-dependent resonance frequency shifts in high quality resonators agrees with predictions by the STM. However, according to STM theory, one also expects that as the power of the radiation applied to resonator is increased, the TLS in the dielectrics become saturated, thereby limiting the maximal power that can be dissipated by photons. This results in a strong electric field dependence of the quality factor ${Q \propto \sqrt{\langle n \rangle}}$, above a critical value $n_c$. Here ${\langle n \rangle \propto {\cal E}^2}$ is the average number of microwave photons within the superconducting resonator and ${\cal E}$ is the electric field applied to the resonator. This power dependence is indeed observed in many resonators characterized by intrinsic loss tangent ${\sim 10^{-3}}$ at very low powers \cite{lindstrom2009properties,pappas2011two}. However resonators characterized by lower intrinsic loss at low powers typically show much weaker power dependence \cite{macha2010losses,khalil2011loss,sage2011study,wisbey2010effect}. The failure of the STM to predict the power dependence of the quality factor for the high quality resonators is an indication of a serious gap in our understanding of TLS in amorphous insulators.

It has been suggested \cite{faoro2012internal} that the anomalously weak power dependence can be explained by the assumption that in high-Q superconducting resonators the TLS located at the interfaces are subject to stronger interactions than the TLS located in the bulk dielectrics studied previously.  These TLS interactions lead to a drift of the TLS energies that results in a logarithmic dependence of their absorption on the radiation power in agreement with the data. 

Further evidence of the importance of TLS interactions was reported in a study of $1/f$ noise in high quality Nb resonator  with Pt capping \cite{burnett2014evidence}. In that work, data was shown that could not be fit by the conventional STM, instead pointing  toward the model in Ref.~\cite{faoro2012internal}  that contains two different types of TLS: ``slow'' classical  fluctuators that can be thermally activated even at millikelvin temperatures with very long time-constants  and ``fast'' coherent TLS with typical energy scales of GHz. Only the latter can directly couple to resonators or qubits, but the two types interact. In particular,  the parameters of the coherent TLS are affected by nearby slow fluctuators. This interaction causes the coherent TLS to move in and out of resonance with a microwave resonator or qubit:  resulting in a ``telegraph''-type signal with the familiar $1/f$ noise spectra. 
It was shown that  all features of the low frequency noise in superconducting resonators are captured by this simple model: namely, the frequency dependence of the spectrum $S_y \sim f^{-1}$, the temperature dependence $S_y \sim T^{-\beta}$ and the applied power dependence ${S_y \sim \langle n \rangle ^{-1/2}}$ as well as the saturation of
the noise with the power at the temperature dependent level \cite{burnett2014evidence,faoro2015interacting} .More recently,  several other groups have published works  related to the effects of interacting TLS on resonators and qubits \cite{ramanayaka2015evidence,lisenfeld2015observation,skacel2015probing,muller2015interacting}. 

Very recently Burin et al. \cite{burin2015low} argued that interactions between TLS might not be so relevant. They calculated the $1/f$ frequency noise by using the usual STM \textcolor{black}{ with the added presence of spectral diffusion} and showed that for amorphous solids characterized by "typical" parameters (namely ${\chi=10^{-3} - 10^{-4}}$, with $\chi=P_0 U_0$ where $P_0$ is the typical density of TLS and $U_0$ denotes the dipole-dipole interaction scale between TLS) in the regime $T<0.1 K$ \textcolor{black}{where the addition of spectral diffusion to the STM predicts an $1/f$ spectrum with  ${S_y \propto T^{-(1+\mu)}}$}, where the additional exponent $\mu$ is associated with the logarithmic temperature dependence of the spectral diffusion width \cite{black1977spectral}. Recently Ramanayaka et al. \cite{ramanayaka2015evidence}  published data that supports this theory in the regime $T<0.1$K.  
However, the experimental data of Burnett et al. \cite{burnett2014evidence}, measured at ${T \sim \hbar \nu_0/k_B}$ where $\nu_0$ is the resonator frequency, cannot be explained by this theoretical result. Burin et al. \cite{burin2015low} therefore argued that the experimental data above $0.1$ K might be explained by assuming that, in the high quality resonators the dimensionless parameter, $\chi$ is much smaller than typical values in amorphous glasses (for example if the density $P_0$ or the interaction $U_0$ between TLS is much smaller than typically expected), also arguing that the relaxation rates of TLS in these resonators can be larger than in ordinary glasses, because of the contribution of conducting electrons in the Pt capping layer. 

In an attempt to  resolve this controversy we here re-analyse some of our previously published data on a platinum capped Nb lumped element resonator and supplement it with data on the loss and the $1/f$ frequency noise in a fractal  $\lambda/2$ Nb resonator without Pt capping.  Prompted by recent experiments by Ramanayaka et al\cite{ramanayaka2015evidence}, we have also studied the ratio between the $1/f$ noise ampitude and the loss tangent.

This paper is organized as follows: we first describe the experimental apparatus and the two different high quality resonators, we then outline briefly the main ideas of our interacting TLS model and recall the formulas for the loss and $1/f$ noise that we derive from the model and we use to fit the data. We present the results and finally discuss implications for the estimates of the parameters characterizing the TLS in these resonators.

\section{Experimental}
\label{Exp}
 
A  dilution refrigerator with a base temperature of 50~mK was used for all measurements and the details of this setup have been described in detail elsewhere\cite{lindstrom2009properties,lindstrom2011pound,burnett2013slow,burnett2014evidence} so only a brief description will be given here. The samples are mounted within a light tight box on a cold stage in contact with the mixing chamber. The box has two microwave lines: one for the input microwave signal which contains 50~dB of attenuation between room temperature and the mixing chamber and room temperature. The second microwave line is for the outgoing microwave signal and contains two microwave 4-8~GHz cryogenic circulators mounted at 700~mK with a HEMT amplifier at 4~K, the noise temperature of this amplifier is $\sim$~4~K.

\begin{figure}[!ht]
	\centering \includegraphics[ 
	width=0.95\columnwidth
	]{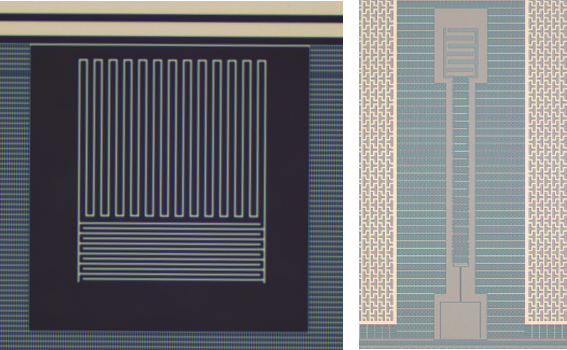}
	\caption{Micrographs of the The two types of resonators used in this work. {\bf left}: Lumped element resonator. { \bf Right:} fractal $\lambda/2$ resonator. }
	\label{fig:resonator}
\end{figure}

Data from two samples are compared in this paper. The first  consists of a 50~nm Nb film, with 5~nm Pt capping layer epitaxially grown\cite{mikhailov1997effect} on a sapphire substrate and patterned into a lumped element (LE) resonator\cite{burnett2014evidence} using photolithography and an SF$_6$/Ar reactive ion etch. The second sample  consists of a 200~nm Nb film sputtered onto a sapphire substrate and patterned into a fractal resonator\cite{de2012magnetic} using electron beam lithography and an SF$_6$/Ar reactive ion etch. Data from both these samples have been previously reported in refs.  \cite{burnett2013pc2} and \cite{burnett2014evidence}. A vector network analyser was used to measure the quality factors of the resonators. The resonance can be fit to an equation of the form $S_{21} = 2\left[2+ \frac{g}{1+2jQ_{l}x}\right]^{-1}$, where $Q_{l}$ is the loaded quality factor, $g$ is the coupling parameter and $x$ is the normalized centre frequency ($x = (\nu_{0}-\nu)/\nu_0$). Measuring the resonator with varying ive allows the power dependent uncoupled loss tangent ($\tan\delta = 1/Q_u$) to be determined.  The \textit{intrinsic} loss tangents  $\tan \delta_i$ were determined by measuring the shift of the centre frequency as a function of temperature \cite{lindstrom2009properties} and fitting to the STM. The parameters for the resonators can be found in table \ref{tab:params}.

\begin{table}
\begin{tabular}{|c|c|c|c|}
\hline 
Sample  & $\nu_{0}$  & $Q_{l}$  & $Q_u$  ( $<n>  \sim 100$)   \tabularnewline
\hline 
\hline 
Nb  & 7.04 GHz  &  24000 & 73000  \tabularnewline
\hline 
Nb+Pt  & 6.68 GHz  & 78000  & 370000  \tabularnewline
\hline 
\end{tabular}
\caption{Device parameters for the resonators used in this work.  $Q_l$ and $Q_u$ denote respectively the loaded and uncoupled quality factor.}
\label{tab:params} 
\end{table}

After the initial characterization and measurement of the loss tangent a Pound frequency-locked loop\cite{lindstrom2011pound} was used to track the frequency jitter in the resonators centre frequency\cite{burnett2013slow,burnett2014evidence}.   This method allows for  high-bandwidth ($\approx$ 10 kHz), high-precision ($\approx$ 1 Hz) direct read-out of the centre frequency of the resonator $\nu_0(t)$ For the data presented here, the frequency jitter was  measured by fixing the microwave drive and temperature for a period of 1.4 (3) hours for the Nb (Nb+Pt) sample. The microwave drive and temperature dependence of the noise is mapped out by repeating the measurement at new combinations of microwave drive and temperature. The fractional frequency spectra $S_{y}$ (defined as $\langle\delta{\nu}(t)\delta{\nu}(t')\rangle/\nu_{o}^{2}$) are determined by calculating the overlapping Allan deviation (ADEV) for the jitter time series\cite{Rubiola}. This allows for efficient screening of the data since any form of drift that could affect the recorded data over this long time-scales (drifts are readily visible in the ADEV). For time scales $t >$~0.01~s the ADEV reveals a $1/f$ frequency noise characterized by a $h_{-1}$ value. For the 1/f noise $S_y = h_{-1}/f$ and we chose a value of 0.1~Hz to analyse the noise in the more familiar form of a power spectral density, $S_{y}(0.1\text{Hz}) = A$.

\section {Model}
Similarly to the STM,   the TLS in our model are described by pseudo-spin operators,
$S$, and are characterized by an uniform distribution of the energy
difference, $E,$ between their ground and excited state.
In the basis of the eigenstates the Hamiltonian has the simple form
$H=E S^{z}$. The ground and first excited state of
the TLS correspond to a quantum superposition of states characterized
by different atomic configurations. Each TLS is characterized by a
dipole moment $\vec{d}_0=\vec{d}_0(\sin\theta S^{x}+\cos\theta S^{z})$,
which is an operator with both diagonal and off-diagonal components.
$\vec{d}_0$ denotes the difference between the dipole moments in
the two different atomic configurations, its magnitude ${d_0=|\vec{d}_0|}$
sets the scale of the dipole moment. $\theta$ relates the eigenstates of the dipole to a superposition
of its states in real space. Because many dipoles have exponentially
small amplitude for tunnelling between different positions in real space,
the parameters $\theta$ and $E$ are assumed to have distribution
$\mathcal{P}(E,\theta)dE \, d\theta\, \sim P_0/\theta dE d\theta$
for small $\theta$ and $P_0$.

In the STM the interaction between different TLS is
essentially of a dipole-dipole nature with an effective strength given
by the dimensionless parameter $\chi=P_0 U_0$ where ${U_0=d_{0}^{2}/\epsilon_h}$, here
$\epsilon_h$ is the dielectric constant of the medium that host the TLS. Straightforward
analysis shows that the same parameter $\chi$ also controls the phonon mean
free path at low temperatures \citep{leggett1991amorphous}. 
Direct measurements give values of $\chi \approx10^{-3} - 10^{-4}$ in bulk amorphous materials.
We argue that in high quality resonators the TLS located at the interfaces are subject to stronger interactions than the TLS located in the bulk dielectrics. Note that a strong interaction between discrete degrees of freedom always decreases the density of states at low energies, i.e.
${\mathcal P}(E)=P_{0}(E/E_{\text{max}})^{\mu}$. For the Coulomb interaction this effect results in a very large suppression of the density of states and the formation of an Efros-Shklovkii pseudogap \cite{efros1975coulomb}. The dipole-dipole interaction is small and would result in logarithmic
corrections to the density of states for point-like TLS. Because a larger than expected interaction implies that the assumption of point-like
defects is probably wrong,  we do not in our model attempt to derive the probability distribution $\mathcal{P}(E,\theta)$ in some microscopic picture but instead assume that  there is a weak power law dependence of the TLS density of states ${\mathcal P}(E)=P_{0}(E/E_{\text{max}})^{\mu}$ with a small parameter $\mu \approx 0.3$ derived from experiments. 

\begin{figure}[h]
\includegraphics[width=3.5in]{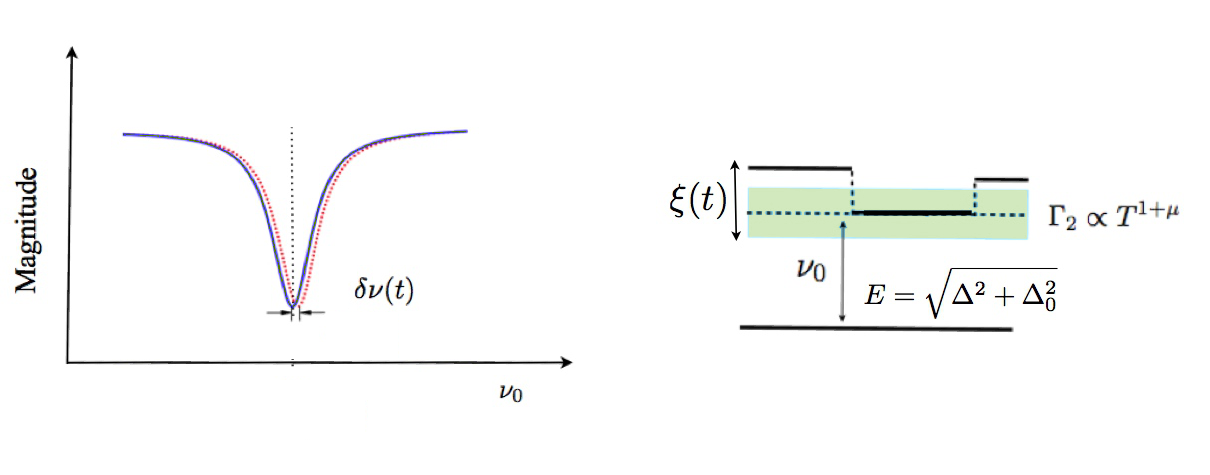} \caption{Schematics of the frequency noise generation in microresonators. The
noise is due to fluctuators that are strongly coupled to resonant
TLS and can induce energy drifts for the resonant TLS larger than
the broadening width $\Gamma_{2}$ by bringing the resonant TLS in
and out of resonance with the resonator. \label{fig:Schematics-of-the-frequency-noise}}
\end{figure}

The calculation of the noise and the loss in the resonators due to an ensemble of interacting disordered quantum TLS is very difficult.  The problem can be simplified  if we distinguish between different TLS: coherent or quantum TLS characterized by a dephasing rate ${\Gamma_2 < E}$ and fluctuators or classical TLS characterized by ${\Gamma_2 \geq E}$. Among the coherent TLS we distinguish between thermally activated TLS with ${E \leq k_B T}$ and resonant TLS having an energy splitting ${E \approx \hbar \nu_o}$ where $\nu_o$ is the frequency of the superconducting resonator. We can then calculate how the relevant physical quantities are affected by the interaction between TLS and we find that both the  $1/f$ noise and the loss at high fields are strongly affected by the switching of classical flucutators that are strongly coupled to resonant TLS. A fluctuator is strongly coupled to a resonant TLS when it is located within a sphere of radius ${R_0=\left (\frac{U_0}{\Gamma_2} \right )^{1/3}}$ centred around the resonant TLS. Because the width $\Gamma_2$ decreases with temperature, the volume of this sphere will grow as the temperature is lowered. Each fluctuator is described as a random telegraph signal with switching rate $\gamma$.  Strongly coupled fluctutators induce an energy drift $\xi(t)$ for the resonant TLS larger than the broadening width $\Gamma_2$ by bringing the resonant TLS in and out of resonance with the resonator (see Fig. \ref{fig:Schematics-of-the-frequency-noise}). The drift $\xi(t)$ is a superposition of the random telegraph signals with a distribution of the switching rates ${{\mathcal P} (\gamma)=P_\gamma/\gamma}$ with normalization constant ${P_\gamma= 1/\ln[\gamma_{max}/\gamma_{min}]}$. 

The loss and the frequency noise are related to the average polarization ${{\bf P}_{\nu_{o}}(t)}$ produced by the
resonant TLS: 
\begin{equation}
{\bf P}_{\nu_{o}}(t)=\frac{1}{2}\langle\vec{d}_{0}\sin\theta\langle S^{+}(t)\rangle_{f}\rangle=\varepsilon_h\chi(\nu_{0},t)\vec{{\cal E}}\label{pol}
\end{equation}
where $\langle\cdot\rangle_{f}$ denotes the average over the distribution of  strongly coupled fluctuators responsible for the energy drift and the average $\langle\cdot\rangle$ is taken over the distribution
of all the coherent TLS and their dipole moments. Here $\vec{{\cal E}}(t)=\vec{{\cal E}}\cos\nu_{0}t$ is the  applied ac electric field.

Specifically, the imaginary part of the average polarization is responsible for the internal quality factor:
\begin{equation}
\frac{1}{Q}=\frac{\int_{V_{h}}{\cal I}m[{\bf P}_{\nu_{o}}(t)]\cdot\vec{{\cal E}}dV}{2\varepsilon \int_{V}|\vec{{\cal E}}|^{2}dV}\label{loss}
\end{equation}
and the relative frequency shift is related to the real part of the average polarization:
\begin{equation}
\frac{\delta\nu(t)}{\nu_{0}}=-\frac{\int_{V_{h}}{\cal R}e[{\bf P}_{\nu_{o}}(t)]\cdot\vec{{\cal E}}dV}{2\varepsilon\int_{V}|\vec{{\cal E}}|^{2}dV}\label{eq:shift}
\end{equation}

The frequency noise spectrum measured in the microresonator is defined  as: 
\begin{equation}
\frac{S_{\delta\nu}}{\nu_{0}^{2}}=\lim_{\tau\to\infty}\frac{1}{\tau}\int_{0}^{\tau}\int_{0}^{\tau}\frac{\langle\delta\nu(t_{1})\delta\nu(t_{2})\rangle}{\nu_{0}^{2}}e^{i\omega(t_{1}-t_{2})}dt_{1}dt_{2}\label{eq:noise}
\end{equation}
 
Calculations carried out in Ref \cite{faoro2012internal,faoro2015interacting}  show that the interaction with strongly coupled classical fluctuators

\begin{itemize}
\item[(i)] results in a  formula for the temperature dependent frequency shift  that agrees with STM theory:
\begin{equation}
\frac{\delta\nu}{\nu_{0}}= \frac{F \tan \delta_i}{\pi} \left [{\cal R}e \Psi \left (\frac{1}{2}-\frac{\hbar \nu_o}{2 j \pi k_B T} \right ) - \log \frac{E_{max}}{2 \pi k_B T} \right ]
\end{equation}
\item[(ii)] does not change the absorption at low powers but changes the square-root dependence of the absorption into a  logarithmic one at high applied fields (we consider the limit of small temperature ${\tanh \frac{\hbar \nu_o}{2 k_B T} \to 1})$:
 \begin{itemize}
 \item[-] at small field: 
 \begin{equation}
\frac{1}{Q_i} =  F \tan \delta_i  \approx F \chi \;
\label{q1}
\end{equation}
\item[]- at large field: 
\begin{eqnarray}
 \frac{1}{Q_u(\cal E)}& =&  P_{\gamma} F \tan \delta_i   \ln \left (\frac{\gamma_{max}}{\Omega} \right ) \; \\
 &\approx&  P_{\gamma} F \chi \ln \left (C\frac{|{\vec{\cal E}}_c|}{|{\vec {\cal E}}|} \right ) \; 
 \label{q2}
 \end{eqnarray}
 \end{itemize}
 where $\gamma_{max}$ is the maximum switching rate of the classical fluctuators coupled to the coherent TLS,  $C$ is a large constant factor and $\displaystyle{\Omega = \frac{|{\vec{\cal E}}|}{|{\vec {\cal E}}_c|} \sqrt{\Gamma_1 \Gamma_2}}$ denotes the Rabi frequency, $\cal{E}_{c}$ is the critical electric field to saturate the TLS. $F$ is a  filling factor which accounts for the fact that the TLS host material volume $V_h$ may only partially fill the resonator volume $V$: 
\begin{equation}
F=\frac{\int_{V_{h}} \varepsilon_h |\vec{{\cal E}}|^2dV}{2\varepsilon \int_{V}|\vec{{\cal E}}|^{2}dV} \approx  \frac{1}{2} \frac{\varepsilon_h}{\varepsilon} \frac{V_h}{V}
\end{equation}
Note that ${\tan \delta_i = \frac{\pi}{3} P_0 \frac{d^2}{\varepsilon_h}  \sim \chi}$.  

\item[(iii)] results in a large $1/f$ noise with amplitude:

\begin{equation}
A_0=\frac{F^2 P_{\gamma}}{\sqrt{1+|{\vec{\cal E}}|^2/|{\vec {\cal E}}_c|^2}} \frac{\chi}{N_{TLS}(T)} \left( \frac{\nu_0}{T}\right)^\mu
\label{eq:A_theory}
\end{equation}

Where $N_{TLS}(T)$ is the number of thermal TLS coupled to the resonator.

\end{itemize}
 
 \section {Results}

\begin{figure}[!h]
	\centering \includegraphics[ 
	width=1\columnwidth
	]{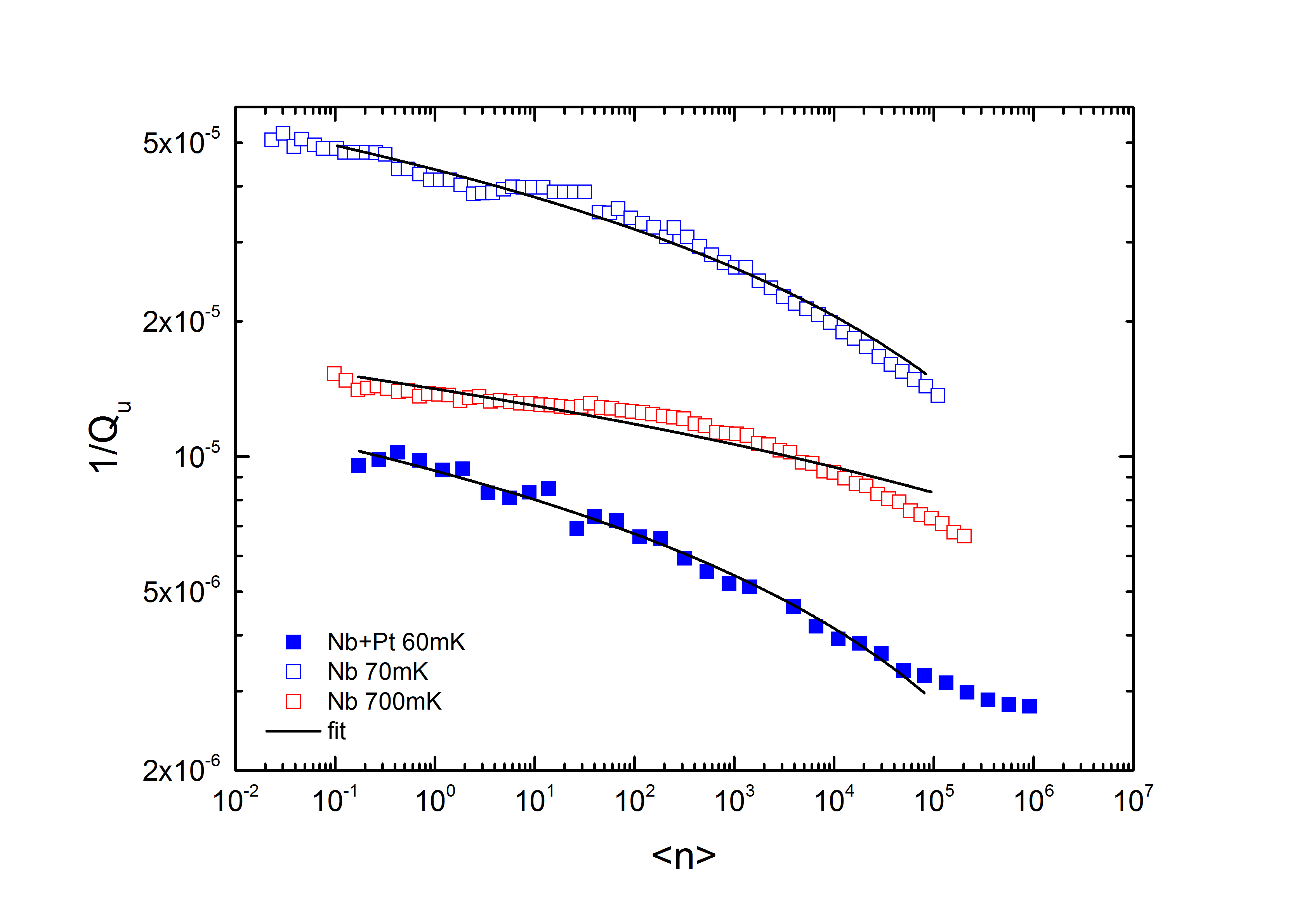}
	\caption{Measurement of the $1/Q_u \propto \tan \delta$. ${\langle n \rangle \propto|{\vec{\cal E}}|^2}$ is the average number of microwave photons within the superconducting resonator. Fit with Eq. (\ref{q2})}
	\label{fig:supfig1} 
\end{figure}

For both the Nb and Nb+Pt resonators the intrinsic quality factors ${Q_i}$  were  first extracted from measurements of the frequency shifts vs. temperature data (this is ${Q_u}$  in the limit of zero field and zero temperature). Then, the power dependent loss $1/Q_u$  was measured with varying microwave drive.  In both the resonators the power dependent loss shows a very weak power dependence (see  figure~\ref{fig:supfig1}) that we fit by using the logarithmic formula given in Eq. (\ref{q2}). 
Data and the prediction of the model are in good agreement.   We note that we do not exclude data from any temperature range when fitting to our model, although we are of course aware that the presence of quasiparticles will have  an effect for temperatures ${T>\hbar \nu_o/k_B}$.
The fact that a logarithmic dependence is found also for the Nb+Pt resonator  implies that the interaction with the conducting electrons present in the Pt capping does not play a role in the relaxation mechanisms of the TLS responsible for the noise.  Note that Eq (\ref{q2}) is calculated in the low temperature limit;  therefore it is unsurprising that the fit is worse for the higher temperature data in Fig~\ref{fig:supfig1}. The values of the intrinsic quality factors extracted from the frequency shifts vs. temperature data and the fits to the loss $1/Q$ are reported in Table \ref{tab:supparams}.

\begin{table}[!h]
\begin{tabular}{|c|c|c|c|c|}
\hline 
Sample  & $\nu_{0}(GHz)$  & $F \chi$ &  $P_\gamma F\chi $  &  $C |{\vec{\cal E}}_c| (V/m)$   \tabularnewline
\hline 
\hline 
Nb  & 7.04 & $1.2\times10^{-5}$ & $1.22 \times 10^{-6}$  &  $3.78 \times 10 ^{7}$ \tabularnewline
\hline 
Nb+Pt  & 6.68  &   $1.1\times10^{-6}$ & $ 2.80 \times 10^{-7}$ &  $4.57 \times 10^{5}$ \tabularnewline
\hline 
\end{tabular}
\caption{Experimental parameters used in the calculations. \textcolor{black}{The data in fig~\ref{fig:supfig1} is fit to equation~\ref{q2} to extract the values for $P_{\gamma}F\chi$ and $C |{\vec{\cal E}}_c|$, while the vales for $F\chi$ are found from a separate measurement of the intrinsic loss tangent.}}
\label{tab:supparams} 
\end{table}

Assuming that the TLS are situated in a surface layer $\approx 10$ nm thick, numerical simulations give a filling factor $F\approx0.01$ for the fractal Nb resonator and ${F < 0.01}$ for the lumped Nb+Pt resonator.  We conclude that in these devices the values of ${\chi \approx10^{-3}}$. 
Notice that we find that the values of the loss tangent estimated from the fit to Eq. (\ref{q2}) at high fields are smaller than the ones obtained in the measurements of the intrinsic loss tangent at zero fields. This is consistent with our model: in fact,  we predict that in the limit of strong applied field the classical fluctuators coupled to resonant TLS cause a  drift of the energy splitting and consequently the additional small contribution  ${P_\gamma = 1/\ln(\gamma_{max}/\gamma_{min})}$ resulting from averaging over the probability of the  switching rates of the classical fluctuators must be taken into account. By examining the data we find  $P_\gamma = 0.1$ for Nb resonator and ${P_\gamma= 2 \times 0.1}$ for Nb+Pt resonator. \textcolor{black}{Studies of $1/f$ charge noise in single-electron transistor and charge qubits report a spectrum that extends from a few Hertz up to a few MHz \cite{kafanov2008charge}, implying $\Gamma_{max}/\Gamma_{min} \approx $~10$^4$ and therefore $P_{\gamma} \approx \text{1}/\ln(\text{10}^{4}) \approx \text{0.1}$}. The fact that $P_\gamma$ is similar for such different resonators provides a further indication that the same mechanism of relaxation are at play in both devices.  

\begin{figure*}
	\centering \includegraphics[width=1.5\columnwidth	]{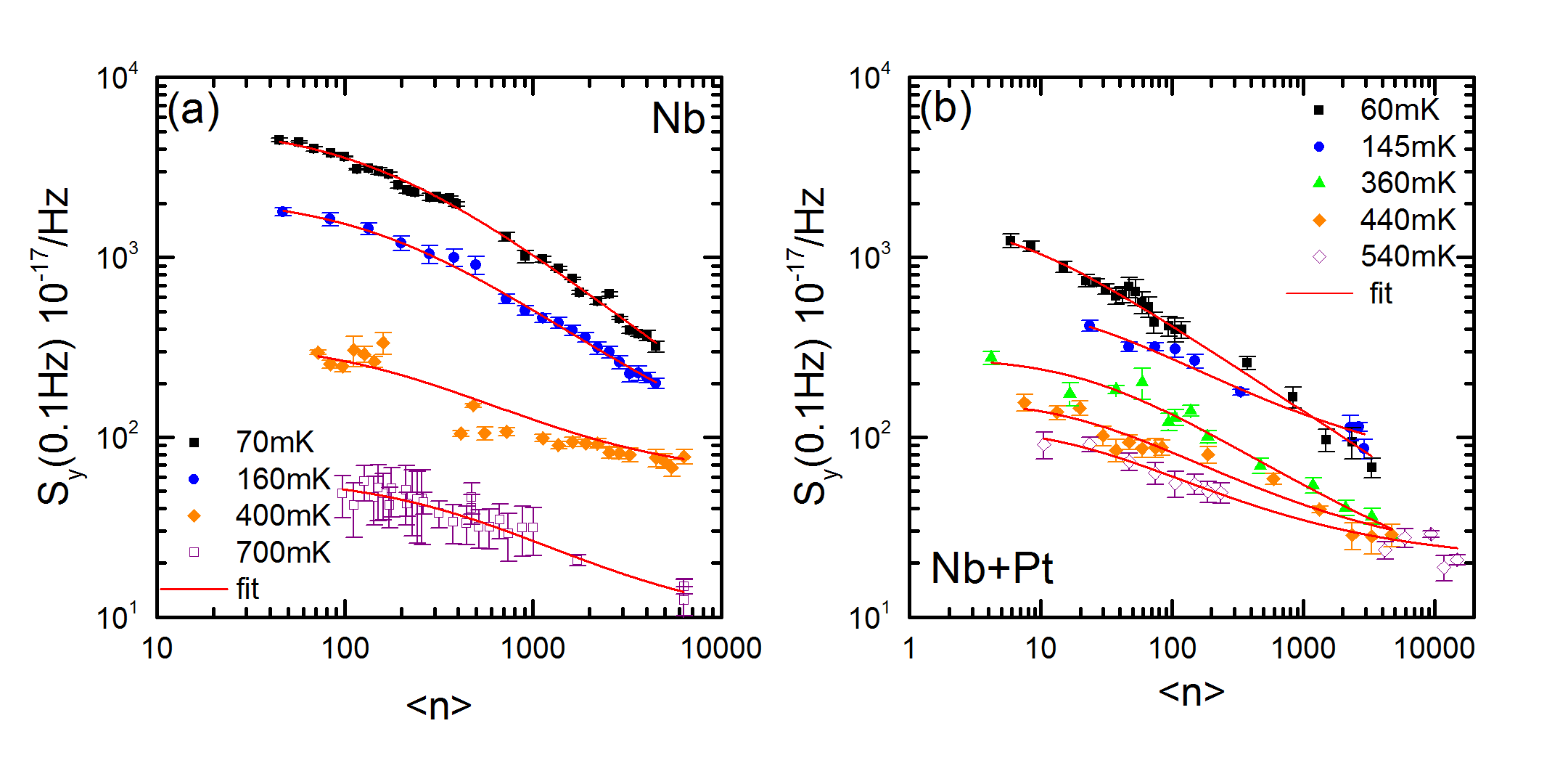}
	\caption{Power spectral density (PSD) of the $1/f$ noise is measured at $S_{y}(\text{0}.1~\text{Hz})$ in varying
		temperature and for different average photon energies in the resonator.
		Shown in red is fit to a power law highlighting an inverse temperature
		dependence. The noise saturates at a power-dependent level above the
		system noise floor of $S_{y}(0.1~\text{Hz})$~=~5x10$^{-17}$. The error
		bars indicate type~A uncertainties. }
	\label{fig:noise} 
\end{figure*}
Fig.~\ref{fig:noise} shows the microwave drive and temperature dependence of the amplitude $A$ of the $1/f$ frequency spectrum in the two resonators. We note that the larger loss tangent in the Nb sample leads to increased sensitivity to temperature fluctuations (due to permittivity shifts induced by thermal excitation of TLS) and consequently the error bars on this data are larger. Improvements to the measurement  setup also made it possible to measure the Nb+Pt resonator at lower microwave drives than the Nb sample. 
From  Eq.~(\ref{eq:A_theory}) we expect a scaling of the amplitude of the noise with the microwave drive as  $A_{0}/(1+\frac{<n>}{n_{c}})^{\beta}$. We fit our data  with $ A_{0}/(1+\frac{<n>}{n_{c}})^{\beta} + C$, where $<n>\propto {\cal E}^2$ is the average number of microwave photons within the superconducting resonator. The values of $A_{0}$ and $n_{c}$ are found to vary with temperature, but we find temperature independent values of $\beta$: 0.5$\pm 0.1$  for the Nb+Pt and  0.8$\pm0.2$ for the fractal Nb sample. Hence, whereas the data for the Nb+Pt sample in good agreement with theory the $\beta$-value from the Nb sample does deviate from the expected 0.5. This could in part be due to that sample mainly having been measured at larger powers where deviations from  eq. \ref{eq:A_theory} are expected; but could also be due to the design since the electric field distribution in the fractal resonator will be less uniform than in the lumped element resonator.  However, the behaviour of both samples is qualitatively the same, despite the design and fabrication process being very different. 

\begin{figure*}
	\centering \includegraphics[ 
	width=1.5\columnwidth
	]{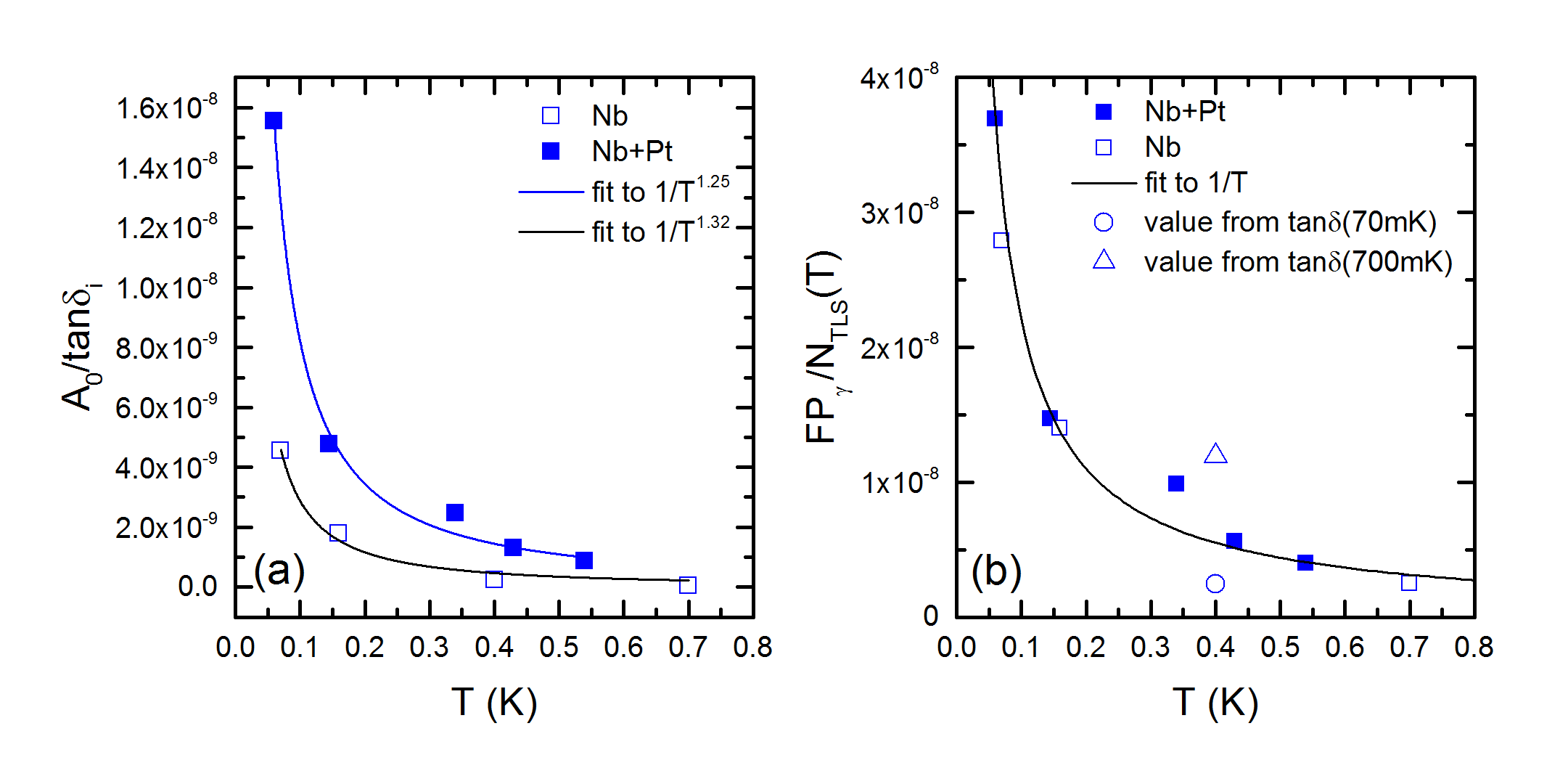}
	\caption{{\bf a)}Temperature dependence of the amplitude of the 1/f noise scaled by the loss tangent for the two resonators. The solid lines represent fits to a dependence $T^{-(1+\mu)}$.  {\bf b)} Temperature dependence of $F P_\gamma/N_{TLS}(T)$ for the  two resonators. The solid lines represent fits to a dependence $T^{-1}$. }
	\label{fig:Nftemp} 
\end{figure*}

In order to examine how the noise amplitude scales with the loss tangent $\tan \delta_i$ we follow  Ramanayaka et al.\cite{ramanayaka2015evidence} and plot the quantity $A_0/\tan \delta_i$ as a function of temperature in Fig.~\ref{fig:Nftemp}a . We find good agreement with  Eq. \ref{eq:A_theory} which predicts a temperature dependence $T^{-(1+\mu)}$ with $\mu = 0.34$ for Nb resonators and ${\mu=0.24}$ for the Nb+Pt capping. However, in our resonators we do not  find that the scaled quantity ${A_0 T/\tan\delta_i}$ is $T$ independent as reported in ref.\cite{ramanayaka2015evidence}. 

To give an order of magnitude estimate of the number of thermal TLS that are indirectly  coupled to the resonator we examine of the ratio of noise to loss. From Eq.~\ref{eq:A_theory} and Eq.~\ref{q2}, we find that in the low field limit the ratio of noise to loss will be $\sim \frac{F P_\gamma }{N_{TLS}(T)} \left(\frac{\nu_{0}}{T} \right)^\mu$. Because the last factor can be estimated using the value of $\mu$ found in Fig~\ref{fig:Nftemp}a, this ratio provides  information on the number of thermally activated TLS, $N_{TLS}(T)$. The resulting ratio  $\sim \frac{F P_\gamma }{N_{TLS}(T)}$ is shown in Fig.~\ref{fig:Nftemp}b. To demonstrate the validity, we show a solid line fit to 1/$T$, which is the expected dependence since $F$ is a geometric parameter and $N_{TLS}$ is expected to scale as $N_{TLS}(T) = P_0 V_h T$. 

We now focus on the Nb+Pt resonator; its lumped  nature makes calculations more straightforward. \textcolor{black}{From Fig.~\ref{fig:Nftemp}b we estimate: ${N_{TLS}(T) \sim 3 \times 10^4}$ at ${T=100 \text {mK}}$. This is equivalent to an average of $\approx 1$ fluctuator/$\mu$m$^2$; comparable to what has been reported for the oxide interface of qubits \cite{martinis2005dec}}.

Assuming the coherent TLS couple to the resonator via their electric dipole moment the relevant volume is approximately given by the area of the interdigitated capacitor (100x200 $\mu \text{m}^2$) multiplied by the thickness $d$ of the layer where the TLS are situated; here we will take $d=10$~nm; giving a total volume of ${V_h \approx 2 \times 10^{-10}~\text{cm}^3}$.  \textcolor{black}{This gives a density of TLS  ${P_0 =N_{TLS}(T)/V_h T \approx \times 10^{-2} \text{nm}^{-3} \text{eV}^{-1}}$, that translates into an interaction scale ${U_0 = \chi/P_0  \approx 0.1 ~\text{eV} ~\text{nm}^3}$ for the intrinsic tangent loss  ${\chi \approx 10^{-3}}$ of the oxide. This is significantly larger than the typical values expected for the phonon strain mediated dipole-dipole interaction between TLS in bulk amorphous material, ${U_0 \approx 10^{-2} \text{eV} ~\text{nm}^3}$ \cite{black1977spectral}). The interaction is related to the dipole moment by ${U_0 \approx 0.1 ~\text{e} ~\text{nm} \approx d_0/\varepsilon_h }$. Assuming $\varepsilon_h \approx 10$ for the oxide, we get ${d_0 \approx 10 \AA \text{e}}$. This is a factor  $\sim 3-4$ larger than values reported for conventional glasses}, ${d_0 = 3 \AA \text{e}}$. The reason for this might be a different microscopic origin of TLS in the surface layer; where they could e.g. be  due to localized electrons. 

Let us now estimate the number of classical fluctuators that are strongly coupled to a resonant TLS. Strongly coupled fluctuators are located within the sphere of radius $\displaystyle{R_0=\left (\frac{U_0}{\Gamma_2} \right )^{1/3}}$ from the resonant TLS. In order to calculate their number, we need first to calculate the width $\Gamma_2$ of the resonant TLS. In the framework of the model \cite{faoro2015interacting}, the width is: ${\Gamma_{2} \sim \chi\ln\left(\frac{\Gamma_{1}^{\text{max}}}{\Gamma_{1}^{\text{min}}}\right)\frac{T{}^{1+\mu}}{E_{\text{max}}^{\mu}}}$, where $\Gamma_{1}^{\text{max}}$ and  $\Gamma_{1}^{\text{max}}$ are the maximum and minimum relaxation rate for the coherent TLS and $E_{max}$ is the maximum TLS level splitting. From spectroscopy of TLS in phase qubits, we estimate ${\ln(\Gamma^{\text{max}}/\Gamma^{\text{min}})\approx 20}$, if we assume that the energy splitting extends to chemical energy scales, i.e. ${E_{max} =100 K}$, we find that ${\Gamma_2 \approx 2 \times 10^{-5} K}$ and the number of strongly coupled fluctuators is $\displaystyle{N_f = \frac{4 \pi}{3} \frac{\chi T}{\Gamma_2}  \approx 20}$, which justifies the assumption  ${N_f \gg 1}$ of the model \cite{faoro2015interacting}.

\textcolor{black}{The data presented in this work and ref. \cite{burnett2014evidence} can be well explained by our model. However, we do note that Burin et al.  have suggested an alternative model and has shown \cite{burin2015low} that this can be made to fit to data presented for the Nb+Pt sample in \cite{burnett2014evidence}; although the fit was restricted to the  ($\langle n \rangle \geq 1$) regime and for data taken at $T>0.1$K. Equivalent data is not present for the Nb sample making a direct comparison to eq. 29 in ref \cite{burin2015low} impossible. Alternatively the data for Nb could be compared to eq. 16 in ref \cite{burin2015low} although this comparison is not attempted as it is non trivial. We do note that the fact that a logarithmic power dependence of the loss is seen in both samples implies that the normal electrons in the Pt capping layer in the NP+Pt sample do not play a role in the relaxation.}. 

\section {Conclusions.}
We have analysed the loss and the low frequency $1/f$ noise in two high quality Nb resonator with and without Pt capping and do not  find any significant difference in the behaviour of the loss and the $1/f$ noise. Both  resonators display similar features in the $1/f$ noise spectrum and a weak logarithmic dependent loss $1/Q$ with varying microwave field. We used the model \cite{burnett2014evidence} to fit the data and find good agreement. We have also studied the ratio between the noise and the loss and extract order of magnitude estimates for the density of states $P_0$, the interaction scale $U_0$ of the thermally activated TLS in the resonators and the number of classical fluctuators that are strongly coupled to a resonant TLS and in our model are ultimately responsible for the noise and the anomalous weak power dependence of the loss of the resonators at high fields. We find a value of ${\chi \approx 10^{-3}}$ in agreement with (but somewhat larger) than the values obtained for amorphous glasses; we also find that the \textcolor{black}{interactions scale ${U_0 \approx 0.1~\text{eV}~\text{nm}^3}$ are a factor $ \sim 10$ larger than the typical value of the phonon strain mediated dipole-dipole interaction between TLS expected in amorphous glasses. This interaction energy corresponds to a dipole moment for the TLS in the oxide layer ${d_0 \approx 10 \AA \text{e}}$ that is again a factor $3-4$ larger than expected for TLS in typical amorphous glasses}.  By comparing the values of the intrinsic quality factors of the resonators extracted from the frequency shifts vs temperature data  (limit of zero field and zero temperature) and the ones evaluated from the measurements of $Q_i$ under varying applied microwave powers in the two different resonators, we exclude the presence of additional  relaxation due to interaction with conduction electrons in the Pt layer and conclude that the mechanism of the noise in the two different resonators is the same.

\section{acknowledgements}
The authors would like to acknowledge  S.~de Graaf, V. L.~Gurtovoi, R.~Shaikhaidarov and V.~Antonov for providing  the samples. We would also like to thank  S.~de Graaf, A. L.~ Burin, J. C.~Fenton, A. Ya. Tzalenchuk and L. Ioffe  for fruitful discussions and comments. 
This work was supported by the NMS, the EPSRC and by ARO W911NF-13-1-0431.

\bibliographystyle{ieeetr}
\bibliography{sust}

\clearpage

\end{document}